\documentclass[cits]{PoS}

\font\tenrsfs=rsfs10 at 12pt
\font\sevenrsfs=rsfs7
\font\fiversfs=rsfs5
\newfam\rsfsfam
\textfont\rsfsfam=\tenrsfs
\scriptfont\rsfsfam=\sevenrsfs
\scriptscriptfont\rsfsfam=\fiversfs
\def\mathscr#1{{\fam\rsfsfam\relax#1}}

\title{Dynamics of WIMPs in the Solar System and Implications for Direct and Indirect Detection}

\ShortTitle{WIMPs in the solar system}

\author{\speaker{Annika H. G. Peter}\\
        Department of Physics, Princeton University, Princeton, NJ 08544\\
	Caltech MS 105-24, Pasadena, CA 91125\\
        \email{apeter@astro.caltech.edu}}

\author{Scott Tremaine\\
        Department of Astrophysical Sciences, Princeton University, Princeton, NJ 08544\\
	School of Natural Sciences, Institute for Advanced Study, Princeton, NJ 08540
        \email{tremaine@ias.edu}}

\abstract{Semi-analytic treatments of the evolution of orbits of weakly
interacting massive particles (WIMPs) in the solar system suggest that the WIMPs bound
to the solar system may enhance the direct detection rate relative to
that of the unbound population by up to a factor of order unity, and
boost the flux of neutrinos from WIMP annihilation in the Earth by up to
two orders of magnitude. To test these important but uncertain
results, we perform a suite of numerical orbit integrations
to explore the properties
of the bound WIMP population as a function of the WIMP mass and the
scattering cross section with baryonic matter.  For regions of WIMP
parameter space presently allowed by experiments, we find that (i) the
bound WIMP population enhances the direct detection rate by at most
$\sim 1\%$ relative to the rate from unbound halo WIMPs; (ii) it is
unlikely that planned km$^3$-scale neutrino telescopes will detect 
neutrinos from WIMP annihilation in the Earth; (iii) the event rate from
neutrinos produced by WIMP annihilation in the Sun may be
much smaller than implied by the usual calculations, which assume that
WIMPs scattered onto bound orbits are rapidly thermalized in the Sun.}

\FullConference{Identification of dark matter 2008\\
		 August 18-22, 2008\\
		 Stockholm, Sweden}

\begin{document}

\newcommand{\sigmapsi}{$\sigma_p^{SI}$}
\newcommand{\sigmansi}{$\sigma_n^{SI}$}
\newcommand{\sigmapsd}{$\sigma_p^{SD}$}
\newcommand{\sigmansd}{$\sigma_n^{SD}$}
\newcommand{\kms}{km~s$^{-1}$}

\section{Introduction}
WIMPs in the solar system may be detected in (i) ``direct detection'' experiments, which measure
the recoil of nuclei during interactions with astrophysical WIMPs \cite{akerib2006b} and have placed interesting constraints on the WIMP
mass $m_\chi$ and spin-independent (or -dependent) elastic scattering cross section with
protons \sigmapsi\ and neutrons \sigmansi\ (\sigmapsd\ and \sigmansd); and (ii) ``indirect
detection'' experiments,  including neutrino telescopes.  The next
generation of neutrino telescopes may detect the
high-energy neutrinos from annihilation of WIMPs captured in the Sun
and Earth \cite{amram1999}.

For a given WIMP model, event rates in direct and indirect detection
experiments are determined by the phase-space distribution function
(DF) of WIMPs.  The fiducial assumption is that
these event rates are dominated by WIMPs from the Galactic halo, passing through the
solar system on unbound orbits \cite{jungman1996}.  However, the following two processes may create additional long-lived populations of WIMPs bound to the solar
system. (i) Damour \& Krauss \cite{damour1999} argued that secular gravitational
interactions with the planets could increase the perihelia of Galactic
WIMPs scattered in the Sun to larger radii, quenching further scattering by solar nuclei and dramatically
increasing the WIMP lifetimes.  Their semi-analytic estimates indicated that
the bound population could increase direct detection rates by up to a factor of $\mathcal{O}(1)$, and enhance the neutrino flux from the Earth by two orders of magnitude \cite{damour1999}.
(ii) Gould and Lundberg \& Edsj{\" o}
\cite{gould1991} estimated that the density of bound WIMPs at Earth
due to gravitational capture and scattering by the planets was $\sim 1\%$ of the density of
unbound WIMPs if additional elastic scattering in the Sun were unimportant.  Because of their low geocentric
speeds the bound WIMPs could enhance the capture rate in the Earth by
up to an order of magnitude, hence boosting the neutrino flux.

Semi-analytic techniques cannot, however, 
describe the full spectrum of behavior in few-body gravitational
systems like the solar system.  Numerical orbit integrations are
the only reliable tool for this task.  In this paper, we summarize the results of a set of such integrations, which
are described further in \cite{peter2008b}.

\section{Simulations}
We assume that the DF of the halo dark matter is
Maxwellian, with a local halo WIMP number density $n_\chi = \rho_\chi / m_\chi$.  $m_\chi$ is the WIMP mass, and we set the local WIMP mass density $\rho_\chi = 0.3\hbox{ GeV cm}^{-3}$.
We set the speed of the Sun relative to the halo to
$v_\odot = 220 \hbox{ km s}^{-1}$, and 
the one-dimensional WIMP velocity dispersion $\sigma = v_\odot /
\surd{2}$.  

The initial conditions for the orbit integration are of two kinds. \emph{Type I:} We employ Monte Carlo methods to determine the positions and
velocities of bound WIMPs after they have scattered elastically
off nuclei in the Sun (using the solar model in 
\cite{bahcall2005}). We run four sets of simulations, each
corresponding to a particular choice of WIMP spin-independent
elastic scattering cross section and mass (\sigmapsi$=10^{-41}$ cm$^2$: $m_\chi = 60$ AMU
; \sigmapsi$=10^{-43}$ cm$^2$: $m_\chi = 60, \hbox{ }150$ and $500$ AMU).  The spin-dependent cross
section is set to zero for simplicity. \emph{Type II:} 
To estimate the DF of WIMPs captured gravitationally, WIMP orbits are followed numerically after they approach within
$1000$ AU of the Sun.  We carry out the simulations with only a
single mass and cross section ($m_\chi = 500$ AMU,
\sigmapsi$=10^{-43}$ cm$^2$), but record the integrated
optical depth from passages through the Sun as a function of time,
which allows us to extend the results to other masses and cross
sections.

Bound WIMPs are followed numerically, admitting the
possibility of additional elastic scattering in the Sun.  We use a simplified solar system containing only one
planet, Jupiter, revolving on a circular orbit about the Sun. The
integration algorithm is symplectic with an
adaptive timestep determined by the heliocentric distance
\cite{mikkola1999}.  We use other methods to integrate orbits when
WIMPs go through the Sun or have close encounters with
planets. Integration is terminated before the age of the solar system if the WIMP escapes from the
solar system or is scattered in the Sun onto an orbit with semi-major axis
$a <0.5$ AU.  Type I simulations followed $10^5$--$10^6$
particles. Type II simulations followed over $10^{10}$ particles, of which $\sim 3\times10^5$ were at
least temporarily bound. 

\section{Results}
\begin{figure}
	\begin{center}
	\includegraphics[width=2.8in,angle=270]{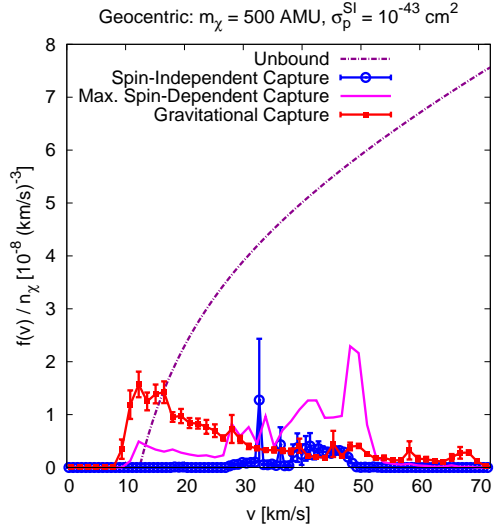}
	\end{center}
	\caption{\label{fig:prl_f1}DFs as a function of 
	geocentric speed $v$, normalized so that 
	$\int f(v)v^2dv$ is the
	number density; $n_\chi$ is the
	halo number density in the solar neighborhood. The dashed
	line is the unbound halo DF.
	The circles represent Type I initial conditions with \sigmapsd$=0$, the squares show the bound DF from gravitational capture
	(Type II), and the solid
	line without points is the estimated DF for Type I capture with \sigmapsd$=
	10^{-36}$ cm$^2$.}
\end{figure}

We present a sample of our results in Fig. \ref{fig:prl_f1}, which shows DFs $f(v)$ at the Earth as a function of geocentric speed $v$, in the case of $m_\chi = 500$ AMU and \sigmapsi$=10^{-43}$ cm$^2$. The dashed line denotes WIMPs from the halo, unbound to the solar system.  Blue circles show the DF of WIMPs with Type I initial conditions with \sigmapsd$=0$.  Red squares represent the WIMPs with Type II initial conditions. Errors are estimated using bootstrap resampling.  We extrapolate our results to obtain a maximum DF due to spin-dependent scattering (the solid magenta line).  The main conclusion in Fig. \ref{fig:prl_f1} is that the total
bound number density is small, no more than $\sim 0.1\%$ of the unbound density.  

We estimate the direct detection rates
of bound WIMPs,
assuming that spin-in\-de\-pen\-dent interactions dominate in the detector
volume.  The analysis thresholds of most
experiments lie above the maximum possible recoil energy for bound
WIMPs ($< 10$ keV).  An exception is the XENON10
experiment \cite{akerib2006b}, for which the
maximum enhancement to the differential event rate is $\sim 1\%$ of
the halo event rate. The enhancement to the total
event rate in the analysis window is $\sim0.1\%$.

We determine event
rates in neutrino telescopes from WIMP annihilation in
the Earth using the DarkSUSY software package, assuming the WIMP is a supersymmetric \cite{gondolo2004}.  We find that the
event rate in an IceCube-sized neutrino telescope \cite{amram1999} is below the detection threshold. 

The Sun is a target for WIMP annihilation searches.  Initial results from the Super-Kamiokande experiment suggest that neutrino telescopes will
be competitive with direct detection experiments in their sensitivity
to \sigmapsd\,, with the standard assumption of rapid thermalization of WIMPs in the Sun \cite{desai2004}.  The neutrino flux from
the Sun is usually estimated assuming that all particles scattered in the
Sun onto bound orbits rapidly reach thermal equilibrium with solar
nuclei, settling into a dense core at the center of the Sun if $m_\chi \gtrsim 4\hbox{ Gev}$ \cite{gaisser1986}. Thermalization is crucial since the
annihilation rate goes as the square of the density.  This picture may
be realistic if $m_\chi \lesssim 1 \hbox{ TeV}$.  However, we find that most WIMPs on Jupiter-crossing orbits are ejected from the
solar system before they can rescatter in the Sun (so long as
\sigmapsd$\lesssim 10^{-38}$ cm$^2$).  There are also long-lived WIMP populations that may not thermalize.  We find that if \sigmapsd$\lesssim 10^{-43}$ cm$^2$ ($\gtrsim 10^{-43}$ cm$^2$) and $m_\chi = 1$ TeV,
20\% (12\%) of WIMPs will never thermalize in the Sun, and rises to
near 100\% (85\%) for $m_\chi = 10$ TeV.

\begin{acknowledgments}
We thank A. Serenelli for providing the standard solar model
in tabular form.  We acknowledge financial support from NASA grants
NNG04GL47G and NNX08AH24G. The simulations were run on
computers at Princeton University supported
by the Departments of Astrophysical Sciences (NSF
AST-0216105) and Physics, and the TIGRESS High
Performance Computing Center. 
\end{acknowledgments}


\begin{thebibliography}{99}


\bibitem{akerib2006b} UK Dark Matter Collaboration, \emph{Phys. Lett.} B \textbf{616}, 17 (2005), D.~S.~Akerib et~al. (CDMS Collaboration), \emph{Phys. Rev.} D \textbf{73}, 011102 (2006); H.~S.~Lee et~al. (KIMS Collaboration), \emph{Phys. Rev. Lett.} \textbf{99}, 091301 (2007); E.~Behnke et~al. (COUPP Collaboration), \emph{Science} \textbf{319}, 933 (2008); J.~Angle et~al. (XENON10 Collaboration), \emph{Phys. Rev. Lett.} \textbf{100}, 021303 (2008); CDMS Collaboration (2008), arXiv:0802.3530.


\bibitem{amram1999} P.~Amram et~al., \emph{Nucl. Phys. B Proc. Suppl.} \textbf{75}, 415 (1999); G.~C.~Hill et~al. (2006), arXiv:astro-ph/0611773; C.~de los Heros et~al. (2008), arXiv:0802.0147.

\bibitem{jungman1996} G.~Jungman, M.~Kamionkowski, and K.~Griest, \emph{Phys. Rep.} \textbf{267}, 195 (1996); J.~Hubisz and P.~Meade, \emph{Phys. Rev.} D  \textbf{71}, 035016 (2005); D.~Hooper and S.~Profumo, \emph{Phys. Rep.} \textbf{453}, 29 (2007).

\bibitem{damour1999} T.~Damour and L.~M.~Krauss, \emph{Phys. Rev.} D \textbf{59}, 063509 (1999);  L.~Bergstr{\"o}m et~al., \emph{JHEP} \textbf{8}, 10 (1999).

\bibitem{gould1991} A.~Gould, \emph{ApJ} \textbf{368}, 610 (1991); J.~Lundberg and J.~Edsj{\"o}, \emph{Phys. Rev.} D \textbf{69}, 123505 (2004).

\bibitem{peter2008b} A.~H.~G.~Peter (2008), in prep.; Ph.D. thesis, Princeton University.

\bibitem{bahcall2005} J.~N.~Bahcall, A.~M.~Serenelli, and S.~Basu, \emph{ApJ} \textbf{621}, L85 (2005).

\bibitem{mikkola1999} S.~Mikkola and K.~Tanikawa,
  \emph{Cel. Mech. Dyn. Astron.} \textbf{74}, 287 (1999);
  M.~Preto and S.~Tremaine, \emph{AJ} \textbf{118}, 2532 (1999).

\bibitem{gondolo2004} P.~Gondolo et~al., \emph{JCAP} \textbf{7}, 8 (2004).

\bibitem{desai2004} S.~Desai et~al., \emph{Phys. Rev.} D \textbf{70},
  083523 (2004).

\bibitem{gaisser1986} T.~K.Gaisser, G.~Steigman, and S.~Tilav, \emph{Phys. Rev.} D \textbf{34}, 2206 (1986).


\end{thebibliography}

\end{document}